\documentstyle[12pt,psfig]{article}
\begin{document}

\begin{center}

{\large \bf Exactly Soluble Model for Nuclear Liquid-Gas Phase Transition\
}

\vspace{1.0cm}

{\bf K.A. Bugaev$^{1,2}$, M.I. Gorenstein$^{1,2}$, 
I.N. Mishustin$^{1,3,4}$ and W. Greiner$^{1}$
}

\end{center}

\vspace{1.0cm}

$^1$ Institut f\"ur Theoretische Physik,
Universit\"at Frankfurt, Germany

\vspace{0.2cm}

$^2$ Bogolyubov Institute for Theoretical Physics,
Kyiv, Ukraine

\vspace{0.2cm}

$^3$ The Kurchatov Institute, Russian Research Center,
Moscow, Russia

\vspace{0.2cm}

$^4$ The Niels Bohr Institute, University of Copenhagen, Denmark

\vspace{2.0cm}

\begin{abstract}
Thermodynamical properties of nuclear matter undergoing multifragmentation
are studied within a simplified version of the statistical model.
An exact analytical solution has been found for the grand canonical ensemble.
Excluded volume effects are taken into account in the thermodynamically self-consistent way. 
In thermodynamic limit the model exhibits a first order liquid-gas phase transition
with specific mixed phase properties.
An extension of the model including the Fisher's term is also studied.
The possibility of the second order phase transition at or above the critical point
is demonstrated.
The fragment mass distributions in the different regions of the phase
diagram are discussed.

\end{abstract}

\vspace{4.cm}


\newpage

{\large \bf I. Introduction}

\vspace{0.2cm}

Nuclear multifragmentation, i.e. multiple production of intermediate
mass fragments, is one of the most spectacular phenomena in
intermediate energy nuclear reactions.
The statistical multifragmentation model (SMM) 
has been developed during last two
decades (see \cite{Bo:95,Gr:97} and references therein).
Numerous comparisons with
experimental data (see e.g. Ref.~\cite{Bot:95,D:96})
show that it is rather successful in explaining many
important features of nuclear 
multifragmentation.
Moreover, there are serious indications \cite{D:96,Po:95}
that multifragmentation in equilibrated systems is
related to a liquid-gas
phase transition in nuclear matter.
Recently a simplified version of the SMM
has been proposed
\cite{Gu:98,Gu:99} to study this relationship.
The calculations within the canonical ensemble of
non-interacting fragments suggest 
the existence of a 1-st order phase transition.
This was demonstrated by increasing the total number
of particles in the system up to A=2800.
These numerical calculations appeared to be rather efficient
due to the recursive formula \cite{Ch:95} which explicitly expressed
the canonical partition function for A particles in terms 
of the canonical partition functions for 1,2,...,A-1 particles.
The recursive procedure essentially simplifies the
numerical evaluations and makes them possible for rather large values
of A. However, the analytical studies of the system behavior
in the thermodynamic limit, i.e. when both the system
mass number A and 
volume V go to infinity, are still missing.
On the other hand, the investigation of the thermodynamic
limit is crucial to proof the existence of a phase transition and
to study its exact nature.

In the present paper we give an exact analytical solution
of the simplified version of the SMM 
performing its complete study in the
thermodynamic limit. 
The accurate treatment
of the excluded volume effects
is an important part of our study. 
We work
in the grand canonical ensemble which
significantly simplifies
the mathematical problems and leads to the explicit analytical
solution in the thermodynamic limit.

The paper is organized as follows. In Sect. II the simplified version of the SMM is formulated.
In Sect. III we introduce the isobaric partition function.
The relationship between its singularities and the phase transition existence
is discussed in Sect. IV. Sect. V is dealing with the first order phase
transition and properties of the mixed phase.
In Sect. VI the possibility of a second order phase transition is demonstrated.
Fragment mass distributions are discussed in Sect. VII.
Sect. VIII is reserved for conclusions.

\vspace{0.5cm}
{\large \bf II. Model Formulation}

\vspace{0.2cm}
Let us consider a system composed of many different nuclear
fragments and characterized by the total number of nucleons $A$,
volume $V$ and temperature $T$.
The system states are specified by the multiplicities $\{n_k\}$
($n_k=0,1,2,...$) of $k$-nucleon fragments ($k=1,2,...,A$).
The partition function of a single fragment with $k$ nucleons is
assumed
to have the standard form \cite{Bo:95}
(here and below the units $\hbar = c = 1$ are used)
\begin{equation} \label{1part}
\omega_k ~=~V~\left( \frac{m T k}{2\pi}\right)^{3/2}~z_k~,
\end{equation}
where $m$ is the nucleon mass and the fragment
mass is
approximated as $m\,k$.
The first two factors in Eq.~(\ref{1part}) originate
from the
non-relativistic thermal motion of the $k$-fragment in the volume $V$
at temperature $T$. The last factor,
 $z_k$, represents the intrinsic partition function of the
$k$-fragment.
For $k=1$ (nucleon) we take $z_1=4$ 
(4 internal
spin-isospin
states) 
and for fragments with $k>1$ we use the expression motivated by the
liquid drop model (see details in Ref.~\cite{Bo:95}): 
\begin{equation} \label{zk}
z_k~=~\exp\left(-~\frac{f_k}{T}\right)~,
\end{equation}
where
\begin{equation} \label{fk}
f_k~=~- [W_{\rm o}~+~
T^2/\epsilon_{\rm o}]~k~+~\sigma (T) k^{2/3}~+~\tau ~T\ln(k)
\end{equation}
is the internal free energy of the $k$-fragment.
Here $W_{\rm o}=16$~MeV is the volume binding energy per nucleon,
$T^2/\epsilon_{\rm o}$ is the contribution of 
the excited states taken in the Fermi-gas
approximation ($\epsilon_{\rm o}=16$~MeV) and $\sigma (T)$ is the
surface free energy tension which is parameterized 
in the following form
\begin{equation} \label{sig}
\sigma (T)~=~\sigma_{\rm o}
~\left(\frac{T_c^2~-~T^2}{T_c^2~+~T^2}\right)^{5/4}\hspace*{-0.2cm}\theta(T_c-T),
\end{equation}
with $\sigma_{\rm o}=18$~MeV and $T_c=18$~MeV.
Finally, $\tau T \ln(k)$ is the phenomenological Fisher's term
\cite{Fi:67}
($\tau$ is a dimensionless constant) which we introduce to generalize our 
discussion.
This form of $z_k$ (with $\tau=0$)
 was used in Refs.~\cite{Gu:98,Gu:99}. 
The symmetry and Coulomb
contributions to the free energy are neglected.
Such a model, however, appears to be a good starting
point for the phase transition studies. 

\vspace{0.5cm}

{\large \bf III. Isobaric Partition Function}

\vspace{0.3cm}
 
The canonical partition function (CPF) of the ideal gas of nuclear
fragments
has the following form:
\begin{equation} \label{Zc}
Z_A(V,T)~=~\sum_{\{n_k\}}~\prod_{k}~\frac{\omega_k^{n_k}}{n_k!}~
\delta(A-\sum_k kn_k)~,
\end{equation}
where $\delta$-function takes care of the total baryon number
conservation.
An important assumption of the model is that the fragments do not overlap
in a coordinate space. This  gives rise to the repulsive interaction which
we take into account in the Van der
Waals excluded volume approximation. This is achieved by substituting
the total volume $V$
in Eq.~(\ref{Zc}) by the free volume $V_f\equiv V-bA$, where
$b=1/\rho_{{\rm o}}$
($\rho_{{\rm o}}=0.16$~fm$^{-3}$ is the normal nuclear density).  
The calculation of the CPF (\ref{Zc})
is difficult because of the restriction $\sum_k kn_k =A$.
This constraint can be avoided by calculating the grand canonical
partition function (GCPF).
Using the standard definition we can write
\begin{eqnarray} 
& &{\cal Z}(V,T,\mu)~\equiv~\sum_{A=0}^{\infty}
\exp\left(\frac{\mu A}{T}\right)~
Z_A(V-bA,T)~\Theta (V~-~bA) \label{Zgc}\\ 
&=&~
\sum_{\{n_k\}}~\prod_{k}~\frac{1}{n_k!}~
\left[\left(V~-~b\sum_k kn_k\right)~\phi_k(T)~\exp(\mu
k/T)\right]^{n_k}~\Theta \left(V~-~b\sum_k kn_k\right)~,\nonumber
\end{eqnarray}
where $\phi_k(T)\equiv \omega_k(T,V)/V$.

The presence of the theta-function in the GCPF (\ref{Zgc}) guarantees
that only configurations with positive value of the free volume 
are counted. However,
similarly to the delta function restriction in Eq.~(\ref{Zc}),
it makes again
the calculation of ${\cal Z}(V,T,\mu)$ (\ref{Zgc}) to be rather
difficult. This problem can be solved \cite{Go:81}
by performing the Laplace
transformation of ${\cal Z}(V,T,\mu)$. This introduces the so-called
isobaric
partition function (IPF):
\begin{eqnarray} \label{Zs}
& &\hat{\cal Z}(s,T,\mu)~\equiv ~\int_0^{\infty}dV~\exp(-sV)
~{\cal Z}(V,T,\mu)\\
&=&~\int_0^{\infty}dV^{\prime}~\exp(-sV^{\prime})
\sum_{\{n_k\}}~\prod_{k}~\frac{1}{n_k!}~\left\{V^{\prime}~\phi_k(T)~
\exp\left[\frac{(\mu  - sbT)k}{T}\right]\right\}^{n_k} \nonumber \\
&=&~\int_0^{\infty}dV^{\prime}~\exp(-sV^{\prime})~
\exp\left\{V^{\prime}\sum_{k=1}^{\infty}\phi_k ~
\exp\left[\frac{(\mu - sbT)k}{T}\right]\right\}~.\nonumber
\end{eqnarray}
After changing the integration variable $V \rightarrow V^{\prime}$,
the constraint of $\Theta$-function has disappeared.
Then all $n_k$ were summed independently leading to the exponential function.
Now the integration over $V^{\prime}$ in Eq.~(\ref{Zs})
can be straightforwardly done resulting in
\begin{equation}\label{Zss} 
\hat{\cal Z}(s,T,\mu)~=~\frac{1}{s~-~{\cal F}(s,T,\mu)}~,
\end{equation}
where
\begin{eqnarray}  
& &{\cal F}(s,T,\mu)~=~ \sum_{k=1}^{\infty}\phi_k ~ 
\exp\left[\frac{(\mu - sbT)k}{T}\right]~\label{Fs}\\
&=&~
 \left( \frac{mT }{2\pi}\right)^{3/2} 
\left[z_1 \exp\left(\frac{\mu-sbT}{T}\right) + \sum_{k=2}^{\infty}
k^{\frac{3}{2}-\tau} \exp\left(
\frac{(\nu - sbT)k -
\sigma k^{2/3}}{T}\right)\right]~. \nonumber
\end{eqnarray}
Here we have introduced the shifted chemical potential
$\nu~\equiv~\mu ~+~W_{\rm o}~+~T^2/\epsilon_{\rm o}$.
Now the GCPT can be obtained by the inverse Laplace transformation
\begin{equation}\label{inv}
{\cal Z}(V,T,\mu)~=~\frac{1}{2\pi i}
\int_{a-i\infty}^{a+i\infty} ds~\exp(sV)~
\hat{\cal Z}(s,T,\mu)~,
\end{equation}
where the integration is taken along the imaginary axis.
In the thermodynamic limit $V\rightarrow \infty$ the system pressure
is defined by the farthest-right singularity, $s^*(T,\mu)$, of  
IPF $\hat{\cal Z}(s,T,\mu)$ (\ref{Zs}) 
(see Ref.~\cite{Go:98} for details):
\begin{equation}\label{ptmu}
p(T,\mu)~\equiv~ T~\lim_{V\rightarrow \infty}\frac{\ln~{\cal Z}(V,T,\mu)}
{V}~=~T~s^*(T,\mu)~.
\end{equation}
The simple connection of the farthest-right $s$-singularity
of $\hat{\cal Z}$, Eq.~(8), to the asymptotic,
 $V\rightarrow\infty$,
behavior of ${\cal Z}$, Eq.~(6), is a general mathematical
property
of the Laplace transform. Due to this property the study of the 
system behavior in the thermodynamic limit
$V\rightarrow \infty$ can be reduced to the investigation of
the singularities of $\hat{\cal Z}$.

\vspace{0.5cm}
{\large \bf IV. Singularities of IPF and Phase Transitions}

\vspace{0.2cm}
The IPF, Eq.~(\ref{Zs}),  has two types of singularities:\\
1). the simple pole singularity
defined by the equation
\begin{equation}\label{pole}
s_g(T,\mu)~=~ {\cal F}(s_g,T,\mu)~,
\end{equation}
2). the essential singularity  of the function ${\cal F}(s,T,\mu)$ 
 it-self at the point $s_l$ where the coefficient 
in linear over $k$ terms in the exponent is equal to zero,
\begin{equation}\label{sl}
s_l(T,\mu)~=~\frac{\nu}{Tb}~.
\end{equation}

The simple pole singularity corresponds to the gaseous phase 
where pressure is determined from the following transcendental
equation
\begin{equation}\label{pgas}
p_g(T,\mu) = \left( \frac{mT }{2\pi}\right)^{3/2} T
\left[z_1 \exp\left(\frac{\mu-bp_g}{T}\right) + \sum_{k=2}^{\infty} 
k^{\frac{3}{2}-\tau} \exp\left(
\frac{(\nu - bp_g)k - 
\sigma k^{2/3}}{T}\right)\right]~.
\end{equation}
The singularity $s_l(T,\mu)$ of the function ${\cal F}(s,T,\mu)$
(\ref{Fs}) defines the liquid pressure
\begin{equation}\label{pl}
p_l(T,\mu)~\equiv~ T~s_l(T,\mu)~=~
\frac{\nu}{b}~.
\end{equation}

In the considered model the liquid phase is represented by an
infinite fragment, i.e. it corresponds to the macroscopic population
of the single mode $k = \infty$. Here one can see the analogy
with the Bose condensation where the  macroscopic population
of a single mode occurs in the momentum space.

In the $(T,\mu)$-regions where $\nu < bp_g(T,\mu)$ the gas phase
dominates ($p_g > p_l$), while  the liquid phase
corresponds to $\nu > b p_g(T,\mu)$. The liquid-gas phase transition
occurs when  two singularities coincide,
i.e. $s_g(T,\mu)=s_l(T,\mu)$.
A schematic  view of singular points is shown 
in Fig.~1a for $T <T_c$, i.e. when $\sigma > 0$.
The two-phase coexistence region is therefore defined by the
equation
\begin{equation}\label{ptr}
p_l(T,\mu)~=~p_g(T,\mu)~,~~~~{\rm i.e.,}~~ \nu~=~b~p_g(T,\mu)~.
\end{equation} 
One can easily see that ${\cal F}(s,T,\mu)$ is monotonously decreasing
function of $s$. 
The necessary condition for the phase
transition is that this function
remains finite in its singular
point \mbox{$s_l=\nu/Tb$:}
\begin{equation}\label{Fss}
{\cal F}(s_l,T,\mu)~<~\infty~.
\end{equation}
The convergence of {\cal F} is determined
by $\tau$ and $\sigma$.
At $\tau=0$ the condition (\ref{Fss}) requires $\sigma(T) >0$.
Otherwise, ${\cal F}(s_l,T,\mu)=\infty$ and the simple pole
singularity $s_g(T,\mu)$ (\ref{pole}) is always the
farthest-right $s$-singularity
of $\hat{\cal Z}$ (\ref{Zs}) (see Fig.~1b). 
At $T>T_c$, where $\sigma(T)=0$, the considered system
can exist only in the one-phase state. It will be shown below
that for $\tau>5/2$  the condition (\ref{Fss})
can be satisfied even at $\sigma(T)=0$.
 
\vspace{0.5cm}
{\large \bf V. 1-st Order Phase Transition and Mixed Phase}

\vspace{0.2cm}

At $T<T_c$ the system undergoes the 1-st order phase transition
across the line $\mu^*=\mu^*(T)$ defined by Eq.(\ref{ptr}).
Its explicit form is given
by the expression ($W\equiv W_{\rm o}+T^2/\epsilon_{\rm o}$):
\begin{equation}\label{muc}
\mu^*(T)~  =~ -~ W~
+~\left(\frac{mT}{2\pi}\right)^{3/2} T b
\left[z_1\exp\left(-~\frac{W}{T}\right) 
 + \sum_{k=2}^{\infty}
k^{\frac{3}{2} -\tau} \exp\left(-~ \frac{\sigma~ k^{2/3}}{T}\right)
\right]~.
\end{equation}
The points on the line
$\mu^*(T)$ correspond to the mixed phase
states. We first consider the case when $\tau=0$.
The line $\mu^*(T)$ (\ref{muc}) for this case is shown in Fig.~2.

The
baryonic density 
is calculated as $(\partial p/\partial \mu)_T$ and
is given by the following formulae in the liquid and gas phases,
respectively: 
\begin{eqnarray}
 \rho_l~ & \equiv & ~
\left(\frac{\partial  p_l}{\partial \mu}\right)_{T}~
= ~ \frac{1}{b}~,\label{rhol} \\
\rho_g & \equiv & ~
\left(\frac{\partial  p_g}{\partial \mu}\right)_{T}~=~
 \frac{ \rho_{id} }{ 1 + b\, \rho_{id} } ~,\label{rhog}
\end{eqnarray}
where the function $ \rho_{id}$ is defined as
\begin{equation}\label{rhoid}
\rho_{id}(T,\mu) = \left( \frac{mT }{2\pi}\right)^{3/2} 
\left[z_1 \exp\left(\frac{\mu-bp_g}{T}\right) + \sum_{k=2}^{\infty}
k^{\frac{5}{2}-\tau} \exp\left(
\frac{(\nu - bp_g)k -
\sigma k^{2/3}}{T}\right)\right]~.
\end{equation}
Due to the condition
(\ref{ptr})
this expression 
is simplified 
in the mixed phase: 
\begin{equation}\label{rhoidmix}
\rho_{id}^{mix}(T)~ \equiv~\rho_{id}(T,\mu^*(T))~=~
\left( \frac{ mT }{2 \pi}\right)^{3/2}
 \left[z_1 \exp\left(-~\frac{W}{T}\right)~+~\sum_{k=2}^{\infty}
k^{\frac{5}{2}- \tau}
\exp\left(-~ \frac{\sigma ~k^{2/3}}{T}\right)\right]~.
\end{equation}
This formula clearly shows that the bulk (free) energy acts in favor of the composite
fragments, but the surface term favors single nucleons.

Since at $\sigma >0$ the sum in Eq.~(\ref{rhoidmix})
converges at any $\tau$, $\rho_{id}$ is finite and according to
Eq.~(\ref{rhog}) $\rho_g<1/b$. Therefore,
the baryonic density has a discontinuity $\Delta\rho =\rho_l-\rho_g >0$
across the line $\mu^*(T)$ (\ref{muc}) shown for $\tau = 0$  in upper panel of Fig.~2. The discontinuities
take place also for
the energy and entropy densities. 
The phase diagram of the system in the $(T,\rho)$-plane 
is shown in lower panel of Fig.~2. 
The line
$\mu^*(T)$ (\ref{muc}) corresponding to the mixed phase
states 
is transformed into
the finite region in the $(T,\rho)$-plane. In this mixed phase
region  of the
phase diagram the baryonic density $\rho$
is a superposition of the liquid and gas baryonic densities:
\begin{equation}\label{mixed}
\rho~=~\lambda~\rho_l~+~(1-\lambda)~\rho_g~.
\end{equation}
Here $\lambda$ ($0<\lambda <1$) is a fraction of the system volume
occupied by the liquid  inside the mixed phase.
Similar linear combination is also valid for the energy density:
\begin{equation}\label{epsmixed}
\varepsilon ~=~\lambda~\varepsilon_l~+~(1-\lambda)~\varepsilon_g~,
\end{equation}
with $(i=l,g)$:
\begin{equation}\label{eps}
\varepsilon_i~\equiv~T\frac{\partial p_i}{\partial T}~+~
\mu\frac{\partial p_i}{\partial \mu}~-~p_i~.
\end{equation}
One finds
\begin{eqnarray}
\hspace*{-0.5cm}&&\varepsilon_l~=~
\frac{T^2/\epsilon_{\rm o}~-~W_{\rm o}}{b}~,\label{epsl}\\
%
%
\hspace*{-0.5cm}&&\varepsilon_g~=~\frac{  1}{1~+~b\rho_{id} } 
\,\,\left\{ \,\,
\frac{3}{2} \,p_g~
+~ \left(    {T^2}/{\epsilon_{\rm o}} -W_{\rm o} \right)\,\rho_{id} 
 \right. \label{epsg}\\
\hspace*{-0.5cm}&&+~\left. {\left(\frac{mT}{2\pi}\right)^{3/2}
\left(\sigma -T\frac{d\sigma}{dT} \right)
\left[ z_1 \, \exp\left(\frac{\mu -bp_g}{T} \right) +\sum_{k=2}^{\infty}
k^{\frac{13}{6}-\tau}\exp\left(\frac{(\nu -bp_g)k-\sigma k^{2/3}}{T}
\,\,\,\right)
\right] }\,\, \right\} \nonumber
\end{eqnarray}

The pressure on the phase transition line $\mu^* (T)$ (\ref{muc}) is
a monotonously
increasing function of $T$
\begin{equation}\label{pcr}
 p^*(T) \equiv p_g(T, \mu^* (T) ) =  
\left( \frac{  m T}{2 \pi} \right)^{3/2} T 
\left[z_1\exp\left(-\frac{W}{T}\right)+
\sum_{k=2}^{\infty}  k^{\frac{3}{2} -\tau}
\exp\left(- ~\frac{\sigma~k^{2/3}}{T}\right)\right]~.
\end{equation}
Fig.~3 shows the pressure isotherms as functions of the reduced density
$\rho/\rho_{\rm o}$  for $\tau=0$.
Inside the mixed phase
the obtained pressure isotherms 
are horizontal straight lines  in accordance with the Gibbs
criterion.
These straight lines go up to infinity when $T\rightarrow T_c-0$.
This formally corresponds to the critical point, $T=T_c$,
$\rho=\rho_c=1/b$ and $p_c=\infty$, 
in the considered case of $\tau=0$. 
For $T>T_c$ the pressure isotherms never enter into the mixed phase region.
Note that, if $\sigma(T)$ would never vanish, the mixed
phase would extend up to infinite temperatures.

Inside the mixed phase at constant density $\rho$ the
parameter $\lambda$ is temperature dependent as shown
in Fig.~4: $\lambda(T)$ drops to zero in the narrow
vicinity of the boundary separating the  mixed phase and 
the pure gaseous phase.
This specific behavior of $\lambda(T)$ causes a strong
increase of the energy density (\ref{epsmixed}) and
as its consequence a narrow peak of the specific heat 
$C_V(T)\equiv (\partial \varepsilon/\partial T)_{\rho}$.
It should be emphasized that the energy density is continuous
at the boundary of the mixed phase and the gas phase,
and the sharpness of the
peak in $C_V$ is entirely due to the strong temperature
dependence
of $\lambda(T)$ near this boundary. 
A narrow peak of the specific heat was
observed in the canonical ensemble calculations   
of Refs.~\cite{Gu:98,Gu:99}. 
However,
in contrast to the expectation in Refs.~\cite{Gu:98,Gu:99},
the height of the $C_V(T)$ peak is not equal to infinity
and its width is not zero in the thermodynamic
limit considered in our study. Note also that the shape
of the $C_V(T)$ depends strongly on the parameter $\tau$ and on the chosen value
of the baryon density
$\rho$.

\vspace{0.5cm}
{\large \bf VI. Possibility of
2-nd Order Phase Transition }

\vspace{0.2cm}

The results presented in Figs.~2-4 are obtained for $\tau=0$.
New possibilities appear
at non-zero values of the parameter $\tau$.  
At $0<\tau \le 5/2$ the qualitative picture remains the same
as discussed above, although there are some
quantitative changes. 
For $\tau > 5/2$  the condition (\ref{Fss}) is also satisfied 
at $T>T_c$ where $\sigma(T)=0$.
Therefore, the liquid-gas phase transition
extends now to all temperatures. Its properties
are, however, different for $\tau >7/2$ and for $\tau <7/2$
(see Fig.~5). 
If $\tau >7/2$ the 
gas density is always lower than $1/b$ as $\rho_{id}$ is finite.
Therefore, the
liquid-gas transition at $T>T_c$ remains
the 1-st order phase transition with discontinuities
of baryonic density, entropy and energy densities.
The pressure isotherms as functions of the reduced density
$\rho/\rho_{\rm o}$ are shown for this case in Fig.~6.

At $5/2 < \tau < 7/2$ the baryonic density of the gas
in the mixed phase, Eqs.~(\ref{rhog},\ref{rhoidmix}),
becomes equal to that of the liquid at $T>T_c$,  since 
$\rho_{id}\rightarrow \infty$ and 
$\rho_g^{mix}=1/b\equiv \rho_{\rm o}$. 
It is easy to prove that the entropy and energy densities
for the liquid and gas phases are also equal to each other.
There are discontinuities only in the derivatives of these densities
over $T$ and $\mu$, i.e., $p(T,\mu)$  has discontinuities
of its second derivatives.
Therefore,  the liquid-gas transition at $T>T_c$ for $5/2 < \tau < 7/2$
becomes the 2-nd order phase transition. 
According to standard definition, the point $T=T_c$,
$\rho = 1/b$ separating the first and second order
transitions is the tricritical point.
One can see that this point is now at a finite pressure.
Fig.~7 shows the pressure isotherms as functions of the reduced density
$\rho/\rho_{\rm o}$.

It is interesting to note that at $\tau >0$ the mixed
phase boundary shown in Fig.5 is not so steep function of $T$ as
in the case $\tau=0$ presented in Fig.3. Therefore, the peak in the specific
heat discussed above becomes less pronounced. 

\vspace{0.5cm}
{\large \bf VII. Fragment Mass Distributions}

\vspace{0.2cm}
The density of fragments with $k$ nucleons 
can be obtained by differentiating
the gas pressure (\ref{pgas}) with respect to the $k$-fragment
chemical potential $\mu_k=k\mu$. This leads to
the fragment mass distribution $P(k)$ in the gas phase
\begin{equation}\label{pkgas}
P_g(k)~=~a_{\rm o}~k^{3/2 -\tau}~\exp(- a_1 k - a_2 k^{2/3})~,
\end{equation}
where $a_1\equiv (bp_g -\nu)/T \ge 0$, $a_2\equiv \sigma/T$
and $a_{\rm o}$ is the normalization constant. 
Since the coefficients $a_{\rm o}, a_1, a_2$ depend on
$T$ and $\mu$ the distribution $P(k)$ (\ref{pkgas})
has different shapes in different points of the phase diagram.
In the mixed phase the condition (\ref{ptr}) leads to
$a_1=0$ and Eq.~(\ref{pkgas}) is transformed into
\begin{equation}\label{pkgasmix}
P_g^{mix}(k)~=~a_{\rm o}~k^{3/2 -\tau}~\exp( - a_2
k^{2/3})~.
\end{equation}
 
The liquid inside the mixed phase is one infinite fragment
which occupies a fraction $\lambda$ of the total system volume.
Therefore, in a large system with $A$ nucleons
in volume $V$ ($A/V=\rho$) the mixed phase
consists of one big fragment with $\lambda V \rho_{\rm o}$
nucleons (liquid) and $(1-\lambda)V\rho_g $ nucleons distributed
in different $k$-fragments according to Eq.~(\ref{pkgasmix}) (gas). 
At low $T$ most nucleons are inside one big liquid-fragment 
with only few small gas-fragments distributed according to
Eq.~(\ref{pkgasmix}) with large $a_2$.
At increasing temperature the fraction of the gas fragments increases
and their mass distribution becomes broader since $a_2(T)$
in Eq.~(\ref{pkgasmix}) decreases.  
Outside the mixed phase region the liquid disappears
and the fragment mass distribution acquires an exponential falloff,
Eq.(\ref{pkgas}). Therefore, the fragment mass distribution is widest
at the boundary of the mixed phase.  
At even higher temperatures, $T>T_c$, the coefficient $a_2$
vanishes. 

Details of the fragment mass distribution depend on the parameter 
$\tau$. At $\tau < 5/2$ we observe a sudden transformation of the large
liquid fragment into light and intermediate mass fragments
in the narrow vicinity of the mixed phase boundary.
This sudden change of the fragment composition has the same origin as a
narrow peak in the specific heat, i.e. a sharp drop of $\lambda(T)$
near the mixed phase boundary (see Fig.~4).
For larger $\tau$ all these changes are getting smoother.

An interesting possibility opens when $5/2<\tau <7/2$.
As shown in Fig.~7 the mixed phase in this case ends at the tricritical
point $T=T_c$, $\rho=\rho_{\rm o}$.
In this point both the coefficients $a_1$ and $a_2$ vanish and the mass
distribution becomes
a pure power law
\begin{equation}\label{power}
P_g(k)~=~a_{\rm o}~k^{ \frac{3}{2} -\tau}~.
\end{equation}
At $\tau >7/2$
the mixed phase exists at all $T$. Thus 
the mass distribution 
of gaseous fragments inside the mixed phase
fulfills a  power-law (\ref{power})at all $T>T_c$.

\vspace{0.5cm}
{\large \bf VIII.  Conclusions}

\vspace{0.2cm}
We have used a simplified version 
of the statistical multifragmentation model (SMM) \cite{Bo:95}
to establish the relationship between
multifragmentation phenomenon and a liquid-gas phase transition 
in nuclear matter. 
Recently, 
in Refs.~\cite{Gu:98,Gu:99}  
interesting peculiarities of this model were found
numerically in the canonical ensemble
formulation. In  present paper this simplified SMM
is solved analytically
by considering the thermodynamic limit $V\rightarrow \infty$
in the grand canonical ensemble.
The progress has been achieved by applying an elegant
mathematical method which reduces the description of phase transitions
to the investigation of  singularities of the isobaric
partition function. 
In this way we have exactly solved the model in the thermodynamic limit.
The excluded volume effects are fully taken into account.

The model clearly demonstrates the 1-st order
phase transition of the liquid-gas type.
It is rather surprising that in thermodynamic limit
the liquid phase emerges as an infinite-mass fragment.
The structure of the mixed phase and some peculiar
properties near its boundary are discussed in details. 
The phase diagram appears to be rather
sensitive to the value of the parameter $\tau$
in the Fisher's free energy term included in our treatment.   
New interesting possibilities for the phase diagram emerge
for $\tau >5/2$  in comparison
with the case when $\tau<5/2$.
The case $5/2<\tau<7/2$ is particularly interesting
because of the appearance of the tricritical point separating the 
1-st and 2-nd order phase transitions.

The results presented in this paper
will be further developed taking into account 
additional physical inputs (e.g. finite size
effects, Coulomb interactions
and symmetry energy) to make the model 
closer to reality.

\vspace{0.5cm}
{\large \bf  Acknowledgments}

\vspace{0.2cm}

The authors gratefully acknowledges
the warm hospitality of the Institute for Theoretical Physics
of the Frankfurt University.
K.A.B. and I.N.M. are grateful to the Alexander von Humboldt Foundation for
the financial support.
M.I.G. acknowledges financial support of DFG, Germany.

\clearpage
\newpage

\begin{figure}

\vspace*{-1.0cm}

\mbox{
\hspace*{1.5cm}\psfig{figure=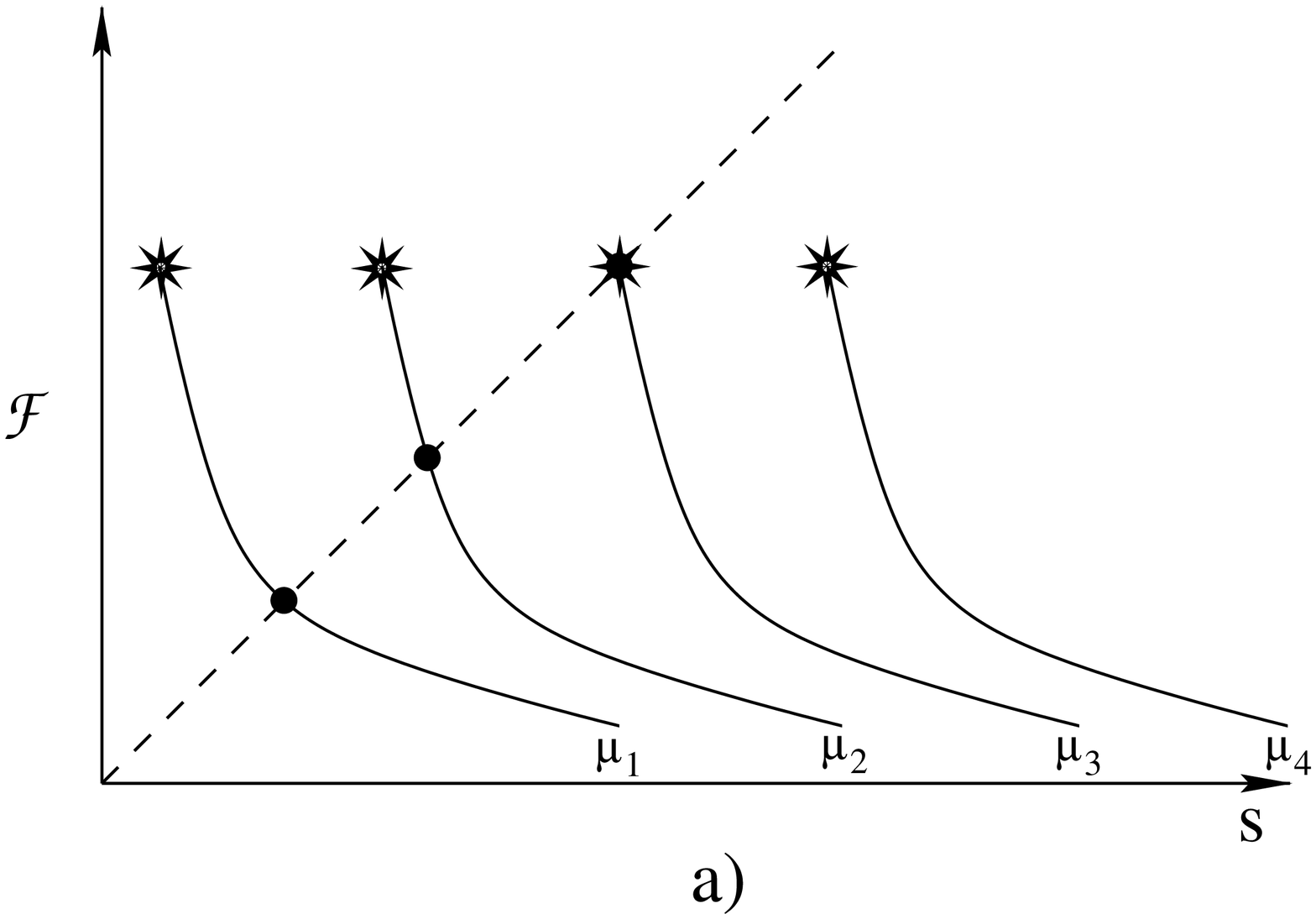,height=9cm,width=12.8cm}
}

\vspace*{1.0cm}

\noindent
\mbox{
\hspace*{1.5cm}\psfig{figure=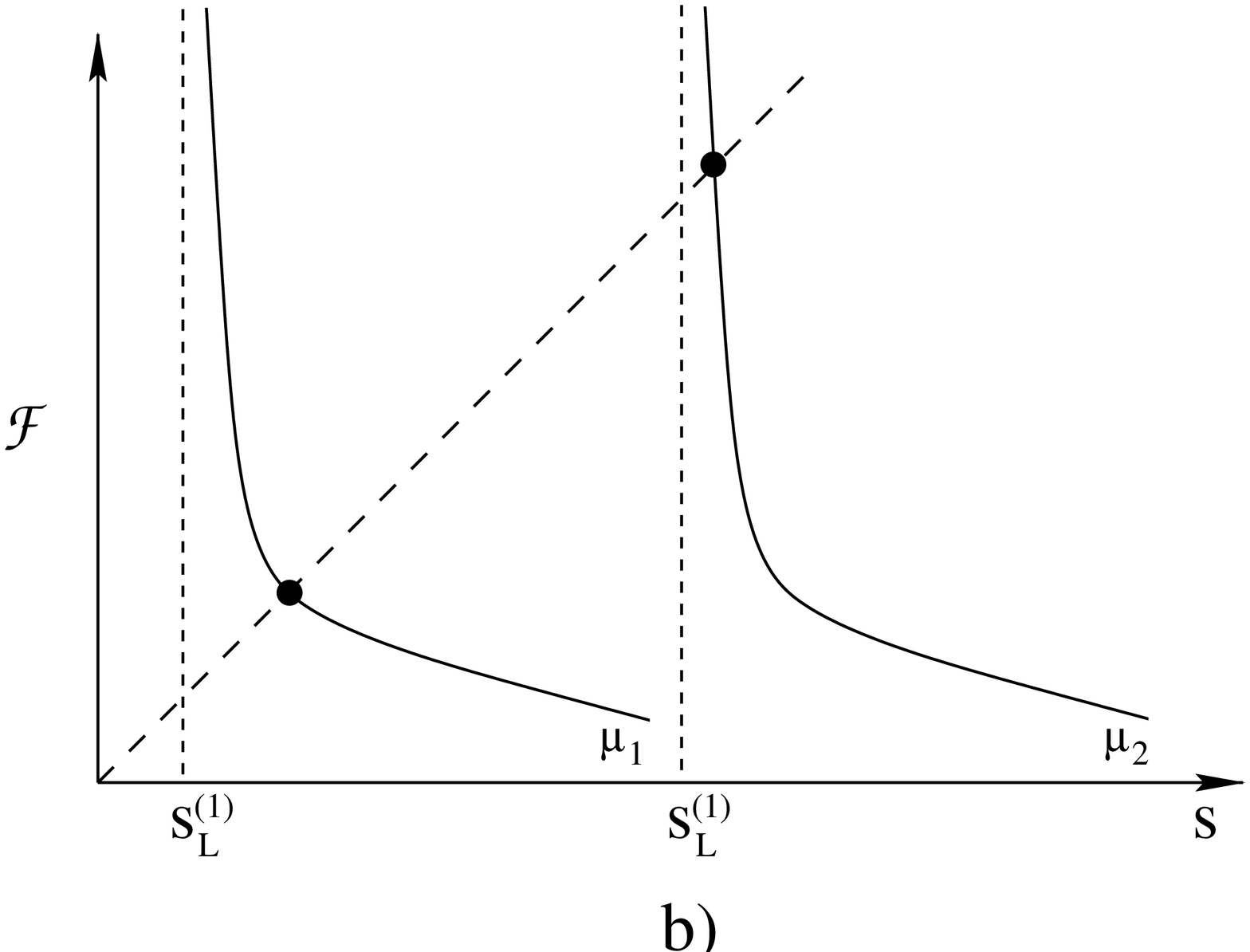,height=9cm,width=12.8cm}
}

\vspace*{1.0cm}

{\bf Fig. 1.}
Schematic view of singular points of the Isobaric Partition Function, Eq. (8),
at $T < T_c$ (a) and $T > T_c$ (b).
Full lines show ${\cal F}(s,T,\mu)$ as a function of $s$ at fixed $T$ and $\mu$, 
$\mu_1 < \mu_2 < \mu_3 < \mu_4$.
Dots and asterisks indicate the simple poles ($s_g$) and the essential singularity of  
function ${\cal F}$ it-self ($s_l$).
At $\mu_3 = \mu^*(T)$ the two singular points coincide signaling a phase transition.

\end{figure}



\clearpage
\newpage
\begin{figure}

\mbox{
\hspace*{3.0cm}\psfig{figure=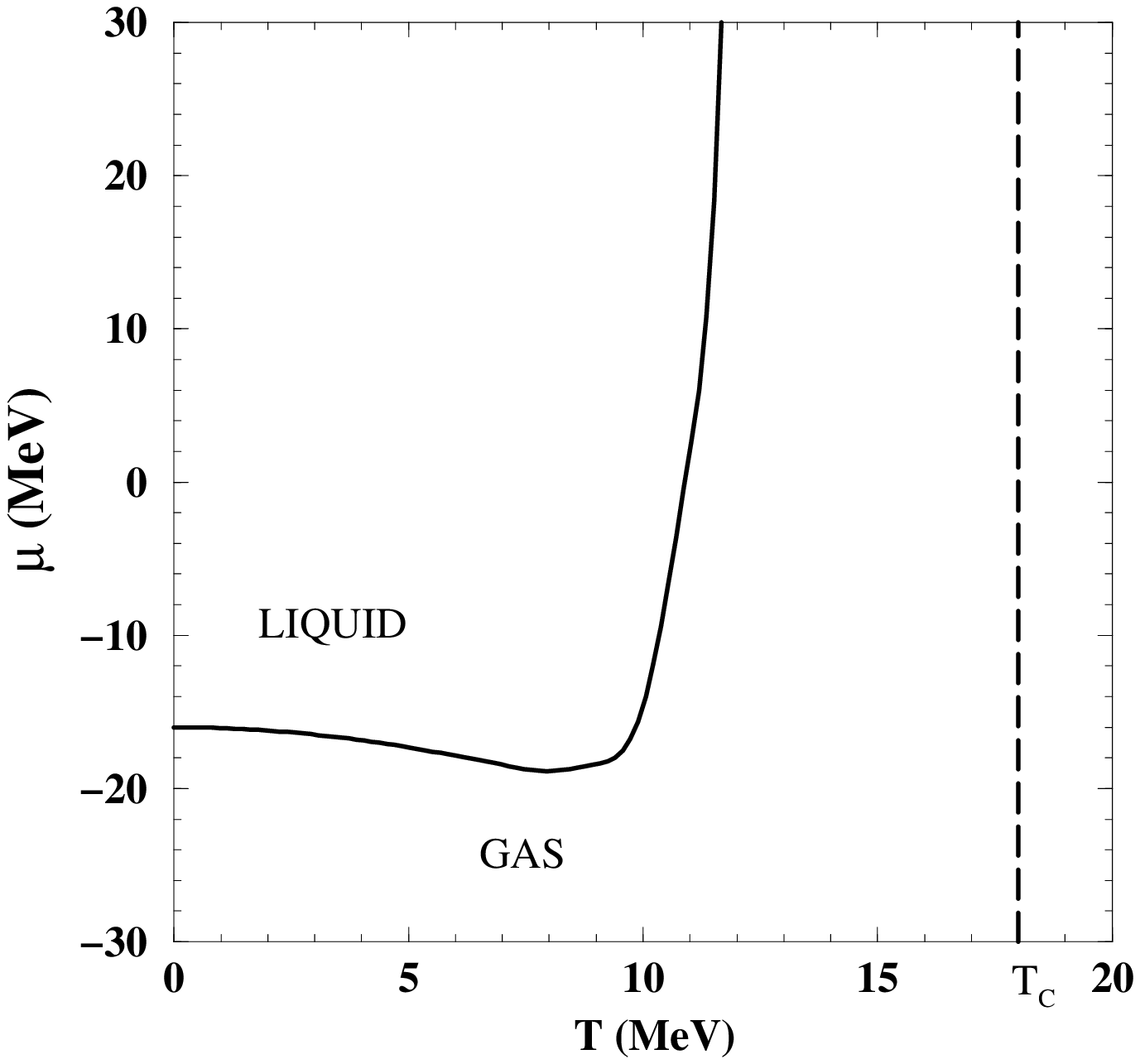,height=9cm,width=9cm}
}

\vspace*{1.0cm}

\noindent
\mbox{
\hspace*{3.0cm}\psfig{figure=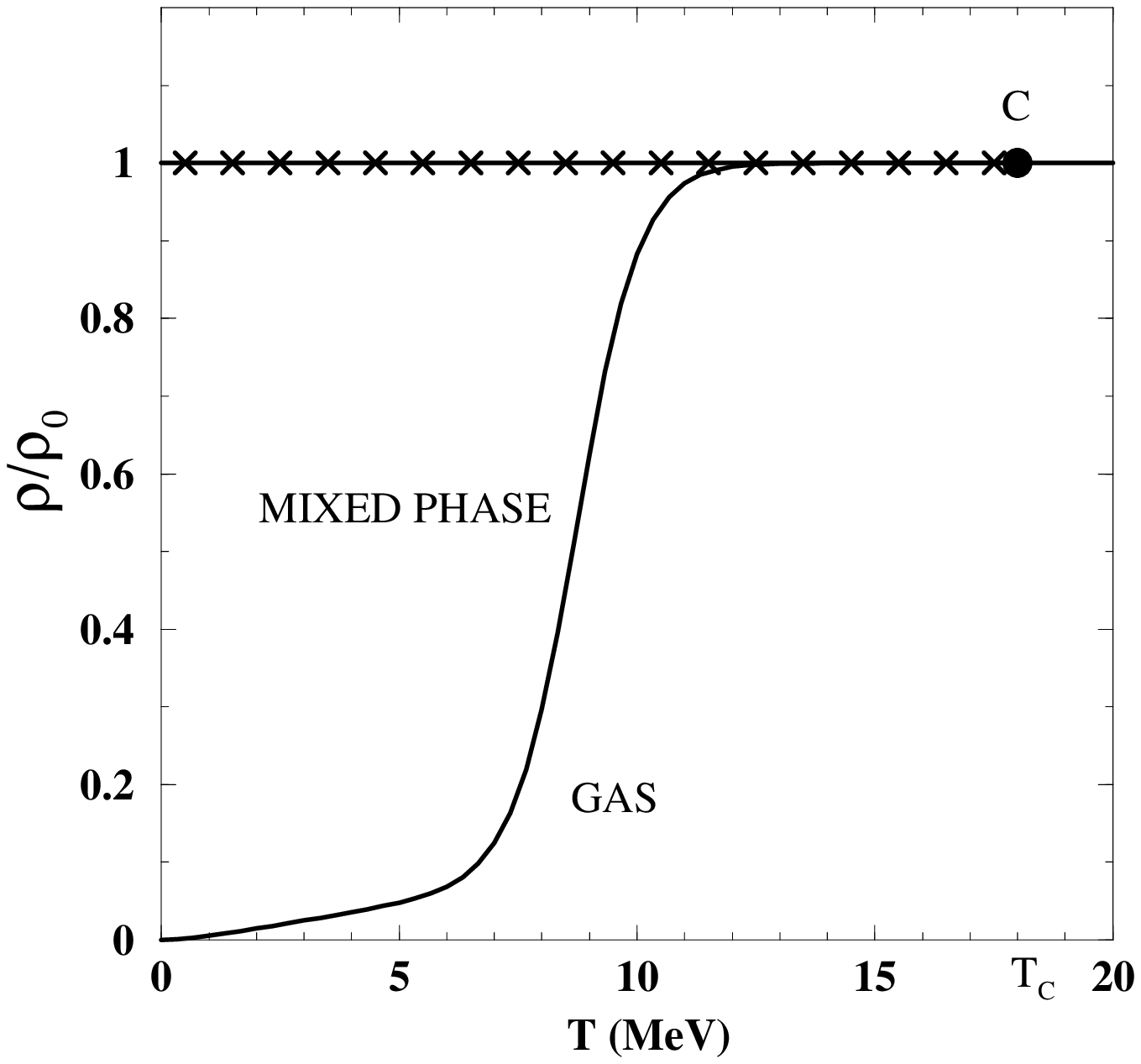,height=9cm,width=9cm}
}

\vspace*{1.0cm}

{\bf Fig. 2.}
Phase diagram in $T-\mu$ (upper panel) and $T-\rho$ (lower panel)
planes for $\tau = 0$.
The mixed phase is represented by the line $\mu^*(T)$ in upper panel and by 
the extended region in the lower panel.
Liquid phase (shown by crosses) exists at density $\rho = \rho_{\rm o}$.
Point $C$ is the critical point.
\end{figure}


\clearpage
\newpage
\begin{figure}

\vspace*{-1.0cm}

\noindent
\mbox{
\hspace*{3.0cm}\psfig{figure=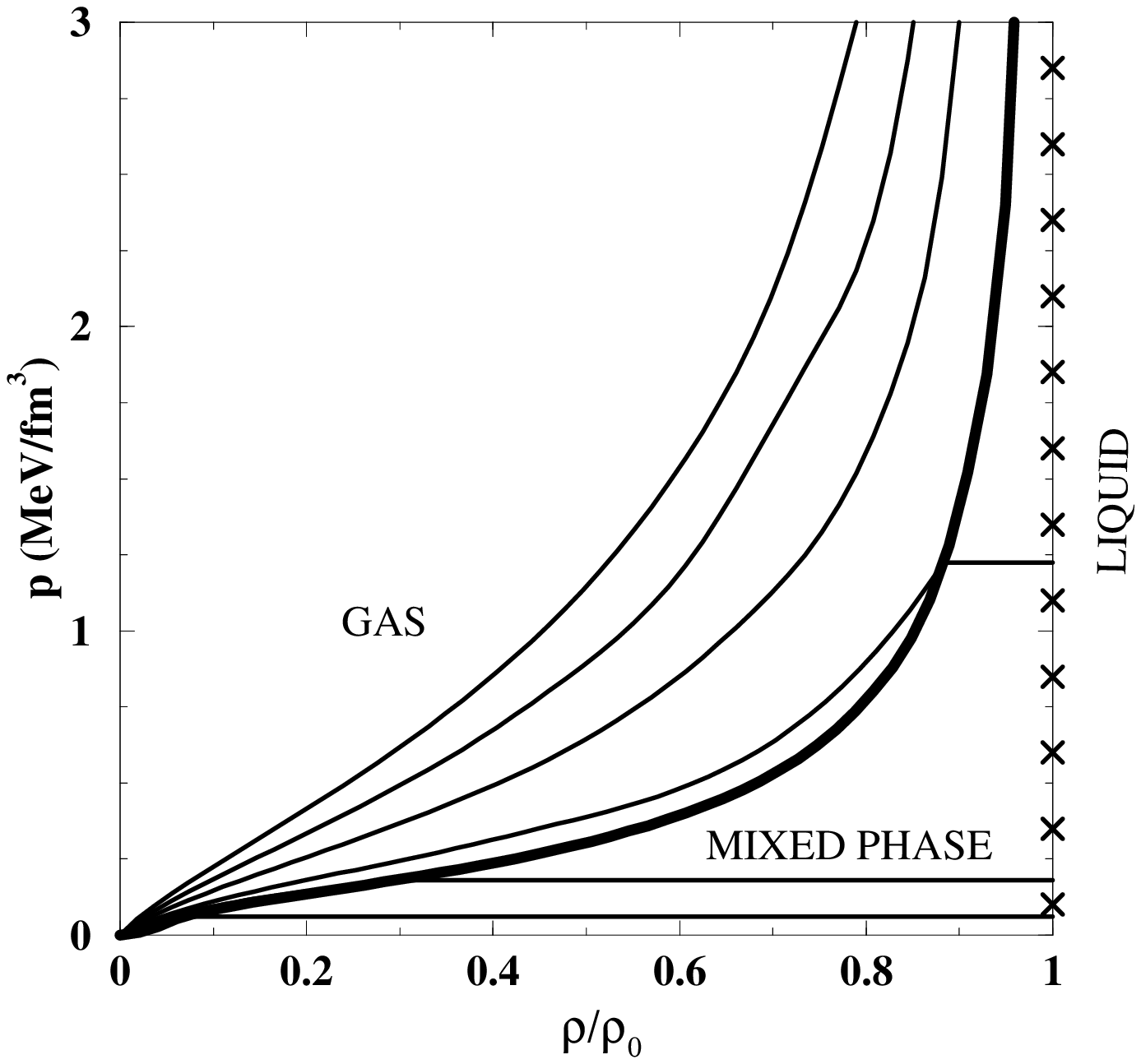,height=9cm,width=9cm}
}

\vspace*{-1.0cm}

{\bf Fig. 3.}
Pressure isotherms (thin solid lines) as functions of reduced density ${\rho}/{\rho_{\rm o}}$
\mbox{for $\tau = 0$.}
The isotherms are shown for $T = 4, 8, 10, 14, 18\,\,{\rm and}\,\, 22$ MeV from bottom to top.
The boundary of the mixed and gaseous phases is shown by the thick solid line. 
Liquid phase is indicated by crosses.
The critical point $T = T_c = 18$ MeV, $\rho = \rho_{\rm o}$ corresponds to infinite pressure.

\vspace*{3.0cm}

\noindent
\mbox{
\hspace*{3.0cm}\psfig{figure=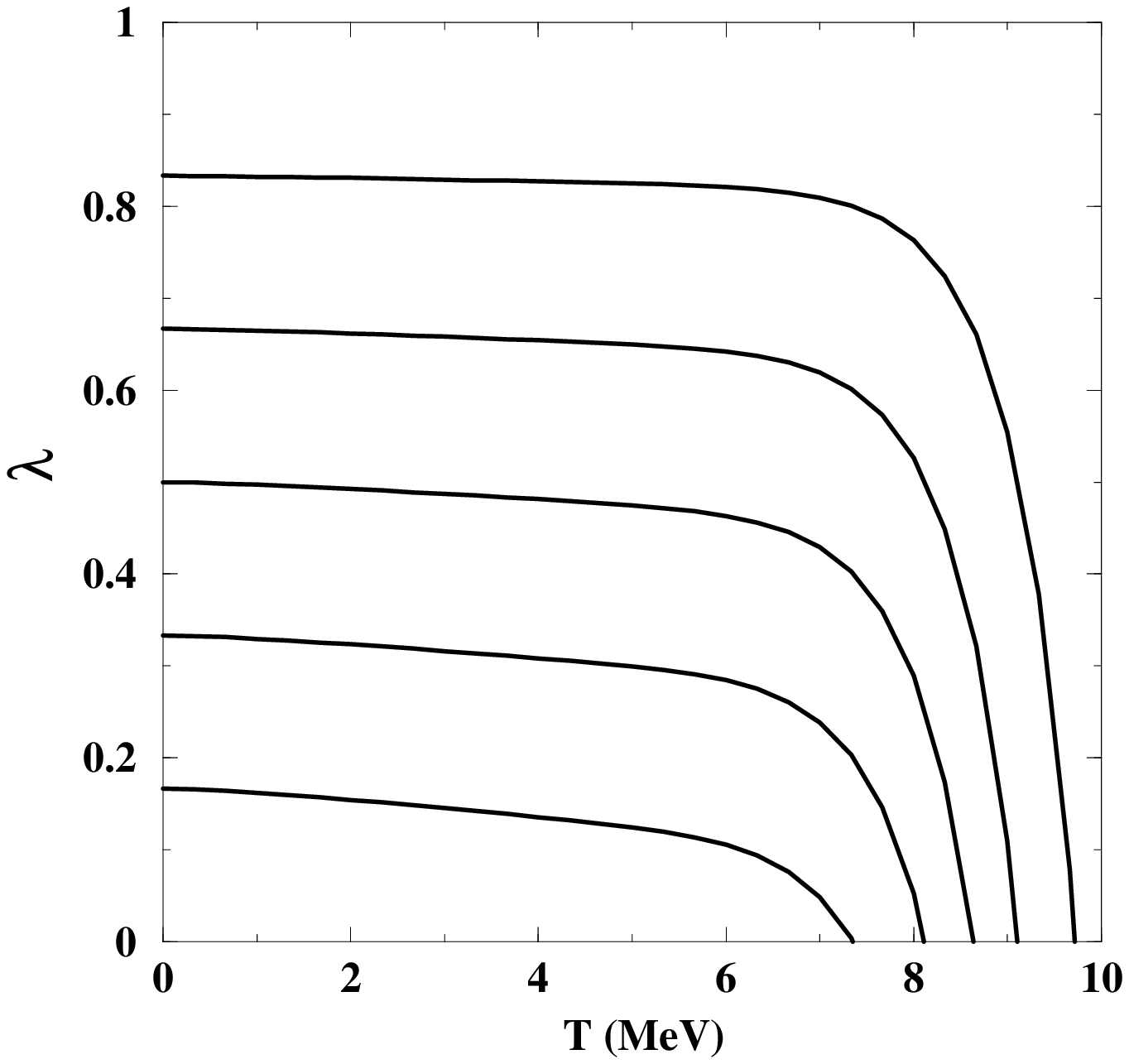,height=9cm,width=9cm}
}

\vspace*{-1.0cm}

{\bf Fig. 4.}
Volume fraction of the liquid phase $\lambda(T)$ for $\tau = 0$
shown as a function of temperature 
at densities ${\rho}/{\rho_{\rm o}} = 1/6, 1/3, 1/2, 2/3, 5/6$ 
(from bottom to top).

\end{figure}

\clearpage
\newpage
\begin{figure}

\vspace*{-1.0cm}

\noindent
\mbox{
\hspace*{3.0cm}\psfig{figure=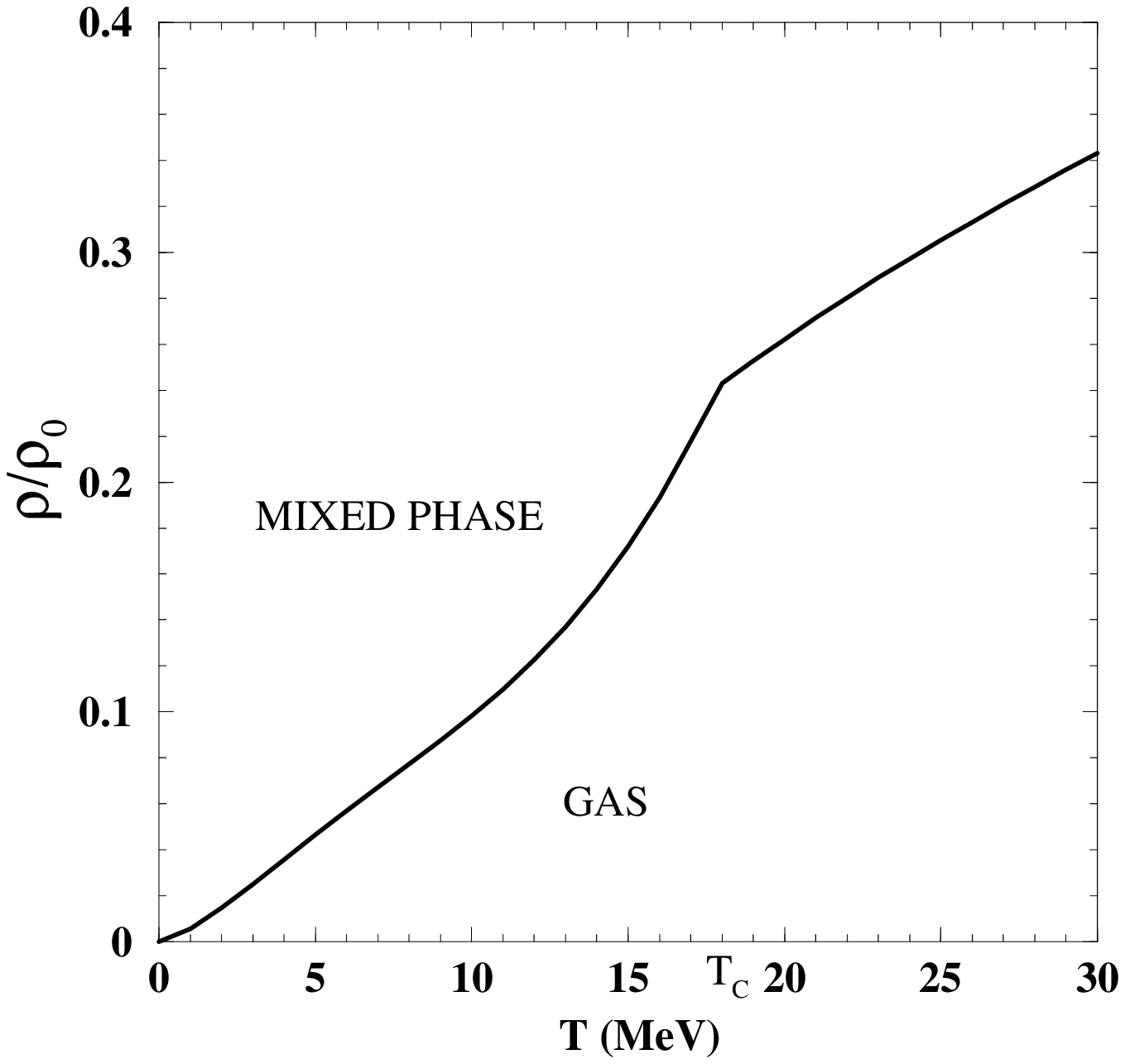,height=9cm,width=9cm}
}

\vspace*{1.0cm}

\noindent
\mbox{
\hspace*{3.0cm}\psfig{figure=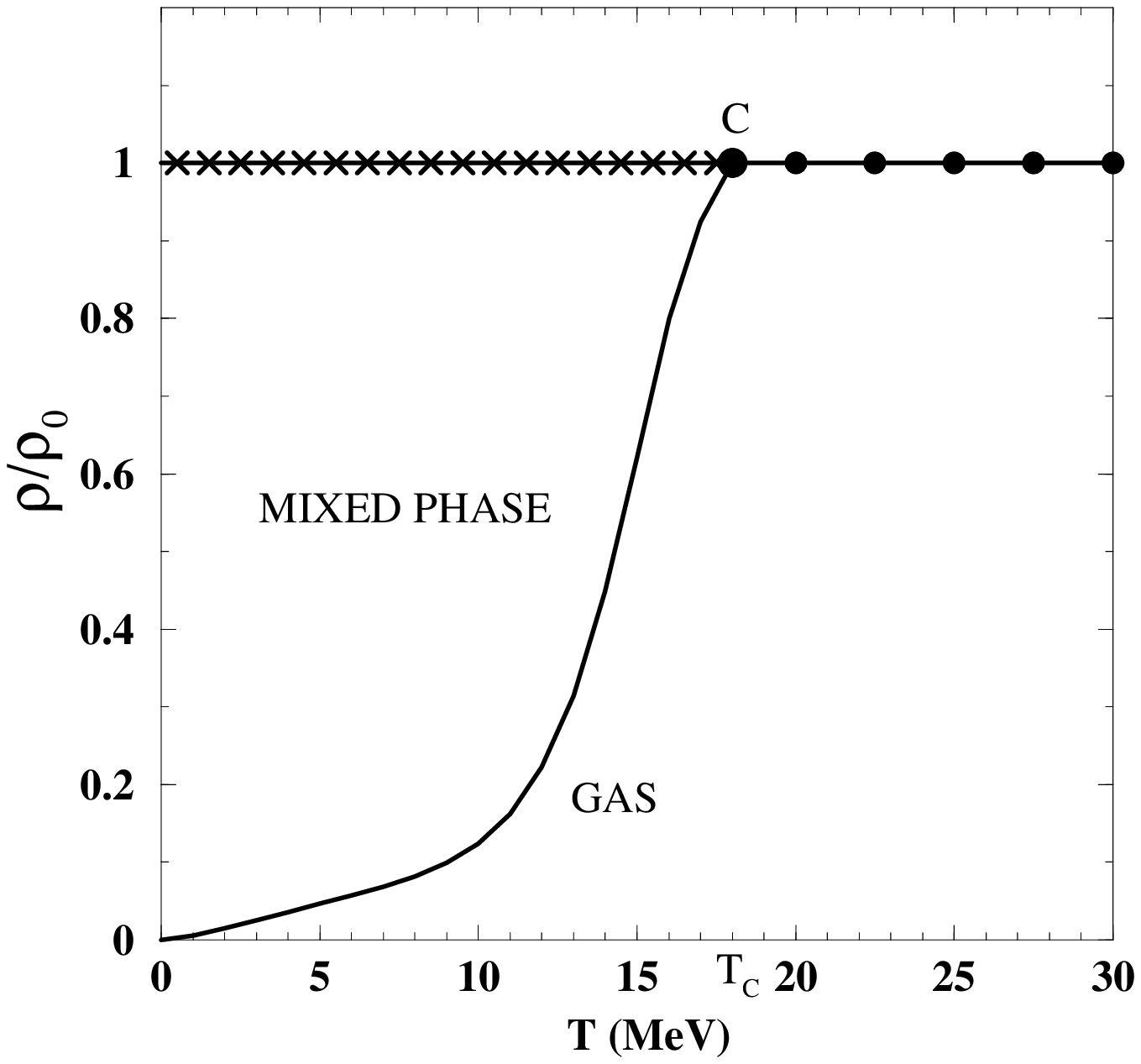,height=9cm,width=9cm}
}

\vspace*{-1.0cm}

{\bf Fig. 5.}
Phase diagrams in 
$T-\rho$ plane for $\tau = 3.6$ (upper panel) and $\tau = 2.6$ (lower panel).
Point $C$ in the lower panel is the tricritical point.
Crosses correspond to the liquid phase of the first order phase transition
and dots correspond to the states of the second order one.

\end{figure}

\clearpage
\newpage
\begin{figure}

\vspace*{-1.0cm}

\noindent
\mbox{
\hspace*{3.0cm}\psfig{figure=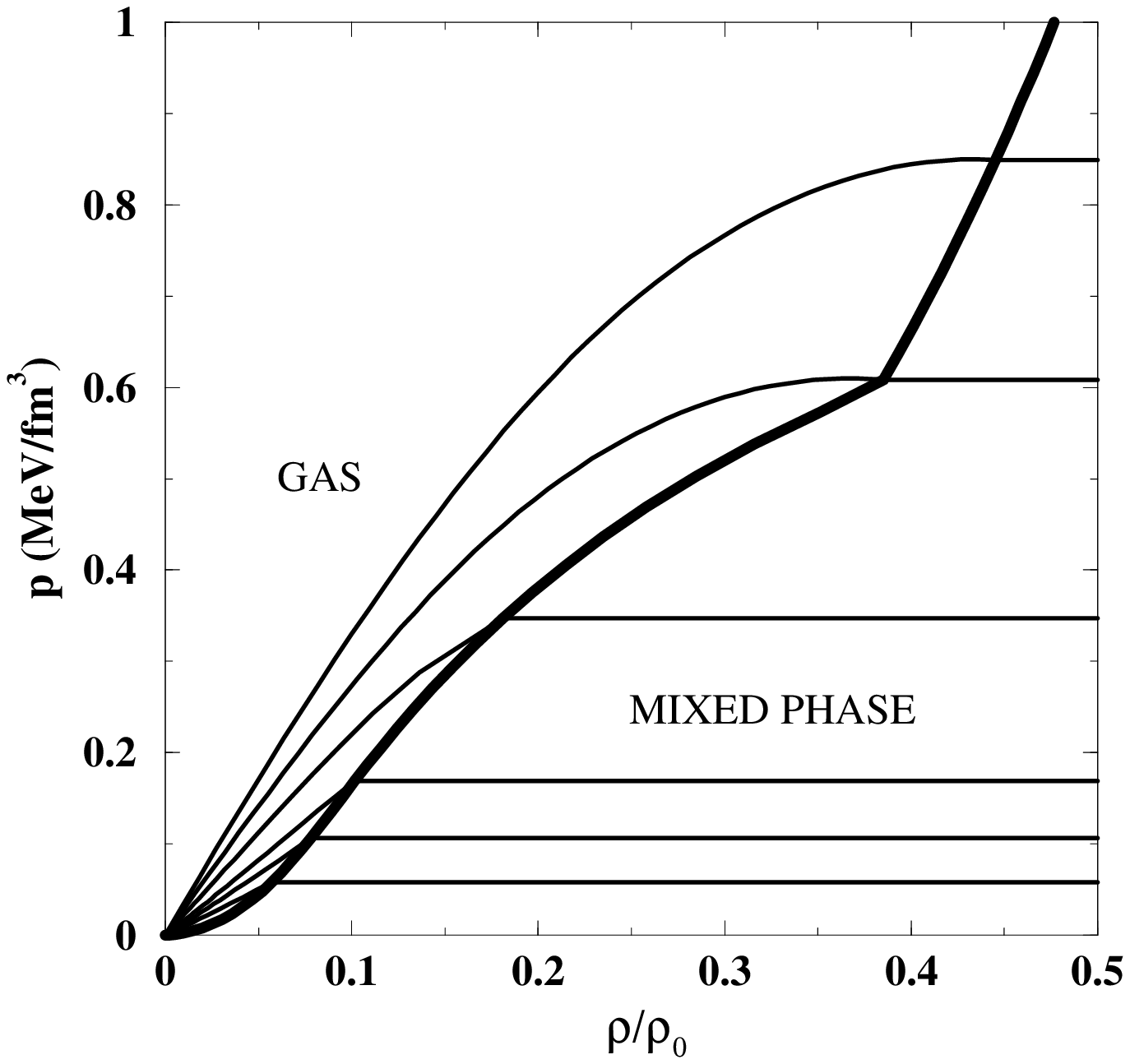,height=9cm,width=9cm}
}

\vspace*{-1.0cm}

{\bf Fig. 6.}
The same as Fig. 3, but for $\tau = 3.6$. 
There is no critical point in this case.

\end{figure}

\clearpage
\newpage
\begin{figure}

\vspace*{-1.0cm}

\noindent
\mbox{
\hspace*{3.0cm}\psfig{figure=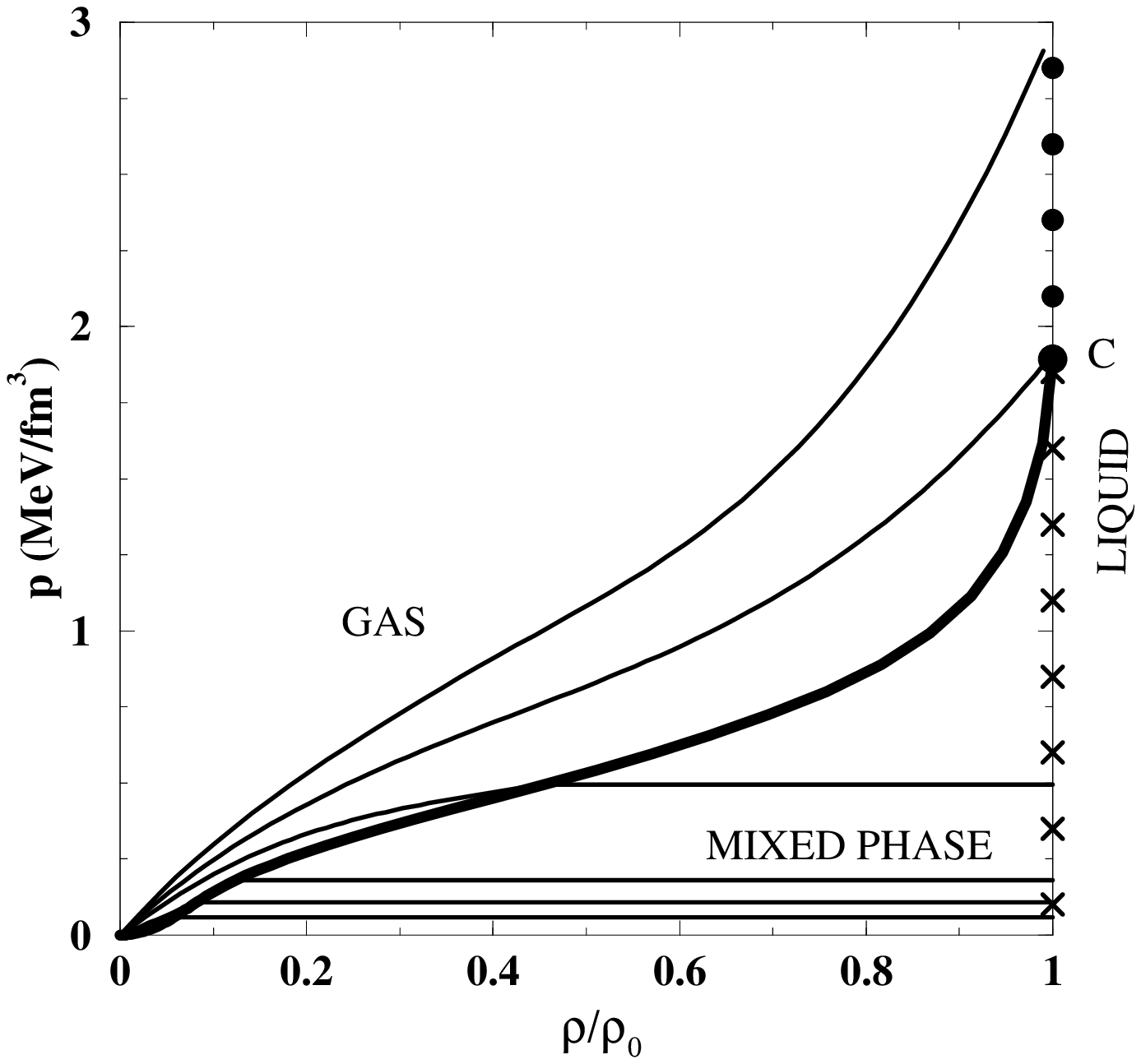,height=9cm,width=9cm}
}

\vspace*{1.0cm}

\noindent
\mbox{
\hspace*{3.0cm}\psfig{figure=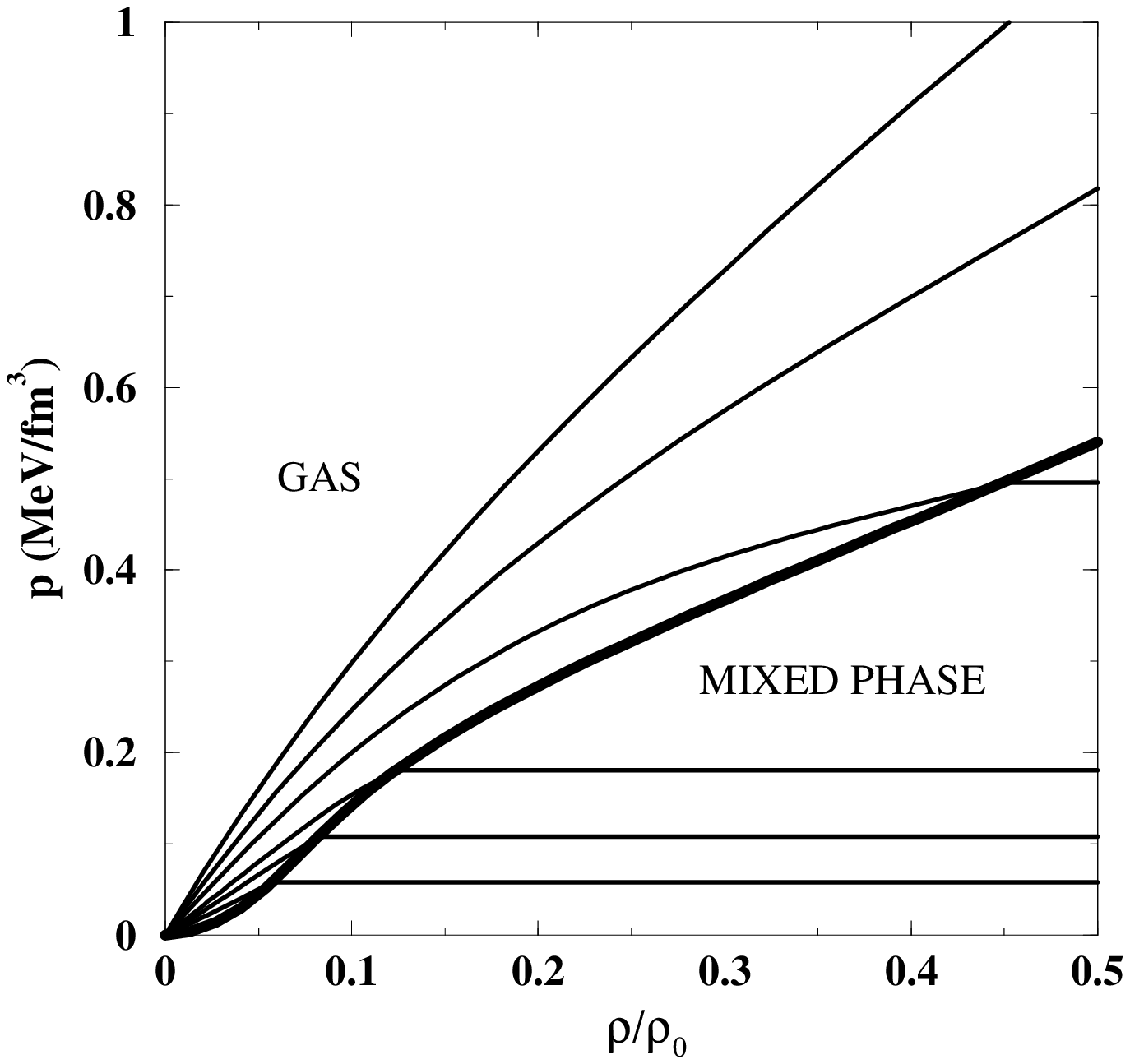,height=9cm,width=9cm}
}

\vspace*{-1.0cm}

{\bf Fig. 7.}
The same as Fig. 3, but for $\tau = 2.6$.
Point $C$ in the upper panel is the tricritical point.
The lower panel differs by the scale. 
Crosses correspond to the liquid phase of the first order phase transition 
and dots correspond to the states of the second order one. 

\end{figure}



\end{document}